# Construction of the UXAR-CT – a User eXperience Questionnaire for Augmented Reality in Corporate Training

Stefan Graser
stefan.graser@hs-rm.de
*Center for Advanced E-Business Studies*
*RheinMain University of Applied Sciences*
Wiesbaden, Germany
ORCID: 0000-0002-5221-2959

Martin Schrepp
martin.schrepp@sap.com
*SAP SE*
Walldorf, Germany
ORCID: 0000-0001-7855-2524

Stephan Böhm
stephan.boehm@hs-rm.de
*Center for Advanced E-Business Studies*
*RheinMain University of Applied Sciences*
Wiesbaden, Germany
ORCID: 0000-0003-3580-1038

***Abstract*—Measuring User Experience (UX) with questionnaires is essential for developing and improving products. However, no domain-specific standardized UX questionnaire exists for Augmented Reality (AR) in Corporate Training (CT). Thus, this study introduces the UXAR-CT questionnaire - an AR-specific UX questionnaire for CT environments. We describe the construction procedure and the evaluation process of the questionnaire. A set of candidate items was constructed, and a larger sample of participants evaluated several AR-based learning scenarios with these items. Based on the results, we performed a Principal Component Analysis (PCA) to identify relevant measurement items for each scale. The three best-fitting items were selected based on the results to form the final questionnaire. The first results regarding scale quality indicate a high level of internal consistency. The final version of the UXAR-CT questionnaire is provided and will be evaluated in further research.**

***Keywords–UXAR-CT; User Experience (UX); UX Measurement; UX Quality Aspects; Questionnaire Construction and Evaluation; Augmented Reality (AR); Corporate Training (CT).***

## I. Introduction

Augmented Reality (AR) enhances the real world with digital content. Therefore, AR refers to three characteristics: (1) a combination of reality and virtuality, (2) real-time interaction, and (3) registration in 3-D [1]. AR can be applied in various domains [2]. Among these, the field of training and education indicates a potential for improving both teaching and learning [3][4]. The field can be divided into academic teaching and Corporate Training (CT). The latter refers to training scenarios in a professional environment. Only a little research was conducted on the latter [2].

Innovative technologies create new interaction paradigms and, thus, a new experience for the user [5]. User Experience (UX) refers to the subjective impression of users towards a product [6]. Measuring the UX is essential for the design and improvement of products. Different methods for measuring the UX can be found in the literature [7]. Applying standardized UX questionnaires to measure subjective impressions is the most established method in UX research.

Various standardized questionnaires can be found in scientific literature [5]. A questionnaire is based on different dimensions, items, and measurement scales, breaking down the construct of UX. However, there is a lack of common ground based on the level of factors and scales [8][9]. Moreover, questionnaires follow different approaches regarding their structure. Not every UX questionnaire can be applied equally in every evaluation scenario. Thus, it is important to use AR-specific UX questionnaires to capture users' perceptions of the technology successfully.

Only three AR-specific UX questionnaires could be found in the literature. However, none of the existing questionnaires refers to CT. This indicates a lack of research concerning UX measurement approaches for AR in CT. Thus, this study introduces the **UXAR-CT**, a domain-specific standardized UX questionnaire for AR in CT. The UXAR-CT was developed on a common concept concerning UX. The UXAR-CT evaluated different AR-based CT scenarios at the Chamber of Crafts for Lower Franconia in Schweinfurt, Bavaria (Germany). This article is based on our previous paper providing initial insights into the questionnaire development and design [10]. In this study, we describe (1) the construction process in more detail, (2) the research design and procedure of the empirical study, and (3) the first evaluation results of the UXAR-CT. In conclusion, the final version of the UXAR-CT questionnaire is proposed.

Section II provides insights into related work regarding the concept of UX and UX questionnaires for AR. Section III describes the construction procedure of the UXAR-CT. Section IV describes the evaluation scenario and the procedure of the empirical study. Evaluation results are shown in Section V. Finally, Section VI gives a conclusion and outlook.

## II. Related Work

This Section II introduces our understanding of the construct UX. We discuss the relevant explanatory approach on which the questionnaire is based. Furthermore, the existing AR-specific UX questionnaires are presented.

### *A. The Concept of User Experience*

UX is proposed as a multidimensional construct with different dimensions regarding the subjective impression of users. Various definitions can be found in the literature. The most common definition is given by ISO 9241-210, which defines UX as a "person's perceptions and responses that result from the use or anticipated use of a product, system or service" [6]. This implies that UX is a subjective construct depending on the user's perception.

Moreover, the definition is rather broad and abstract. Thus, this does not help quantify and measure UX. Different approaches were conducted to break down the construct of UX and achieve a better understanding. One common distinction was made by [11] dividing into *pragmatic* and *hedonic* qualities. Furthermore, some research aimed to consolidate UX factors based on the empirical as well as the semantic similarity on the level of measurement items [8][9][12]–[14].

Schrepp et al. [14] broke down UX into a set of semantically clearly described quality aspects. In this regard, the terms UX factor and UX quality aspect are considered the same. In summary, [14] proposed 16 UX quality aspects shown in the following table I:

TABLE I. CONSOLIDATED UX QUALITY ASPECTS BASED ON [14]

| | UX Quality Aspect | Definition |
|---|---|---|
| (1) | Perspicuity (PE) | Is it easy to get familiar with the product and to learn how to use it? |
| (2) | Efficiency (EF) | Can users solve their tasks without unnecessary effort? Does the product react fast? |
| (3) | Dependability (DE) | Does the user feel in control of the interaction? Does the product react predictably and consistently to user commands? |
| (4) | Usefulness (US) | Does using the product bring advantages to the user? Does using the product save time and effort? |
| (5) | Intuitive Use (IU) | Can the product be used immediately without any training or help? |
| (6) | Adaptability (AD) | Can the product be adapted to personal preferences or personal working styles? |
| (7) | Novelty (NO) | Is the design of the product creative? Does it catch the interest of users? |
| (8) | Stimulation (ST) | Is it exciting and motivating to use the product? Is it fun to use? |
| (9) | Clarity (CL) | Does the user interface of the product look ordered, tidy, and clear? |
| (10) | Quality of Content (QC) | Is the information provided by the product always actual and of good quality? |
| (11) | Immersion (IM) | Does the user forget time and sink completely into the interaction with the product? |
| (12) | Aesthetics (AE) | Does the product look beautiful and appealing? |
| (13) | Identity (ID) | Does the product help the user to socialize and to present themselves positively to other people? |
| (14) | Loyalty (LO) | Do people stick with the product even if there are alternative products for the same task? |
| (15) | Trust (TR) | Do users think that their data is in safe hands and not misused to harm them? |
| (16) | Value (VA) | Does the product design look professional and of high quality? |

This distinction based on UX quality aspects is a common perspective in UX research [14]–[16]. For instance, common UX questionnaires are based on this approach [17]–[22]. Identifying the relevant UX quality aspects that differ depending on the application field and evaluation object is essential. Not every quality aspect suits every evaluation scenario. Thus, the importance of the UX quality aspects regarding the objective must be considered [10][14]. In the following, the existing UX AR questionnaires are described.

### B. UX Questionnaires for Augmented Reality

As described, many UX questionnaires are available in the literature [5]. However, most of them are general and unrelated to AR. Only three AR-specific UX questionnaires were identified, as shown in the following.

1) **Handheld Augmented Reality Usability Scale (HARUS)** [23][24]
2) **Augmented Reality Immersion (ARI) Questionnaire** [25]
3) **Customizable Interactions Questionnaire (CIQ)** [26]

The Handheld Augmented Reality Usability Scale (HARUS) developed by [23][24] specifically focuses on the usability of handheld AR devices. The questionnaire consists of the factors *Manipulability* referring to the ease of handling, and *Comprehensibility* referring to the ease of understanding. Each factor contains eight items. The evaluation is based on a seven-point Likert scale resulting in a computed score from 0 to 100, similar to the System Usability Scale (SUS) [23][24][27].

The Augmented Reality Immersion (ARI) Questionnaire developed by [25] consists of three main scales *Engagement*, *Engrossment*, and *Total Immersion*, and further six subscales *Interest*, *Usability*, *Emotional Attachment*, *Focus of Attention*, *Presence*, and *Flow* with a total of 21 items on a seven-point Likert scale. The focus is on measuring the immersion which concerns the cognitive and emotional absorption. The questionnaire is intended to be applied in location-aware AR settings.

The Customizable Interactions Questionnaire (CIQ) developed by [26] aims to gather the interaction quality between the user and virtual objects in AR scenarios. The questionnaire is based on the five scales *Quality of Interactions*, *Comfort*, *Assessment of Task Performance*, *Consistency with Expectation*, and *Quality of the Sensory Enhancements* with a total of 17 items on a five-point Likert scale [26].

To sum up, the HARUS specializes in AR devices with a specific Usability focus. In contrast, the ARI questionnaire focuses on immersion in location-based settings. The CIQ, in turn, focuses on the quality of interaction. This shows the still great heterogeneity of the questionnaires even in a specific domain.

In summary, the AR-specific questionnaires consist of different UX factors, indicating different measurement focuses. Furthermore, the questionnaires did not refer to CT. It remains unclear which UX aspects are relevant for AR in CT. This indicates a lack of applicability for UX evaluation in the field of CT.

## III. Construction of the UXAR-CT Questionnaire

In the following, the construction of the **UXAR-CT** questionnaire is described. The developement process consists of three steps illustrated in the following:

1) Determination of relevant UX quality aspects (Section III-A, [10])
2) Construction of UX measurement items (Section III-B, [10])
3) Evaluation and derivation of the final questionnaire (Sections IV and V)

Steps (1) and (2) are shortly introduced in Sections III-A and III-B. Detailed information for the initial development phase can be found in the previous paper [10]. Step (3) was conducted within this study. We conducted an empirical study described in Section IV. Based on this, the evaluation results and construction of the final questionnaire are illustrated in Section V.

Finally, we would like to add a statement regarding the previous study. Within this, we declared the UXAR-CT as a measurement approach concerning **Mobile** Augmented Reality. However, several reasons have emerged that cause us to depart from this terminology. On the one hand, the distinction between the terms is still not clearly outlined in the literature. On the other hand, handheld and head-mounted devices were used at the Chamber of Crafts in Schweinfurt. To keep it simple, we will only refer to AR now. This also does not affect the first study on the importance of UX quality aspects. Thus, this change is therefore not considered critical. In the following, the three steps of construction are illustrated.

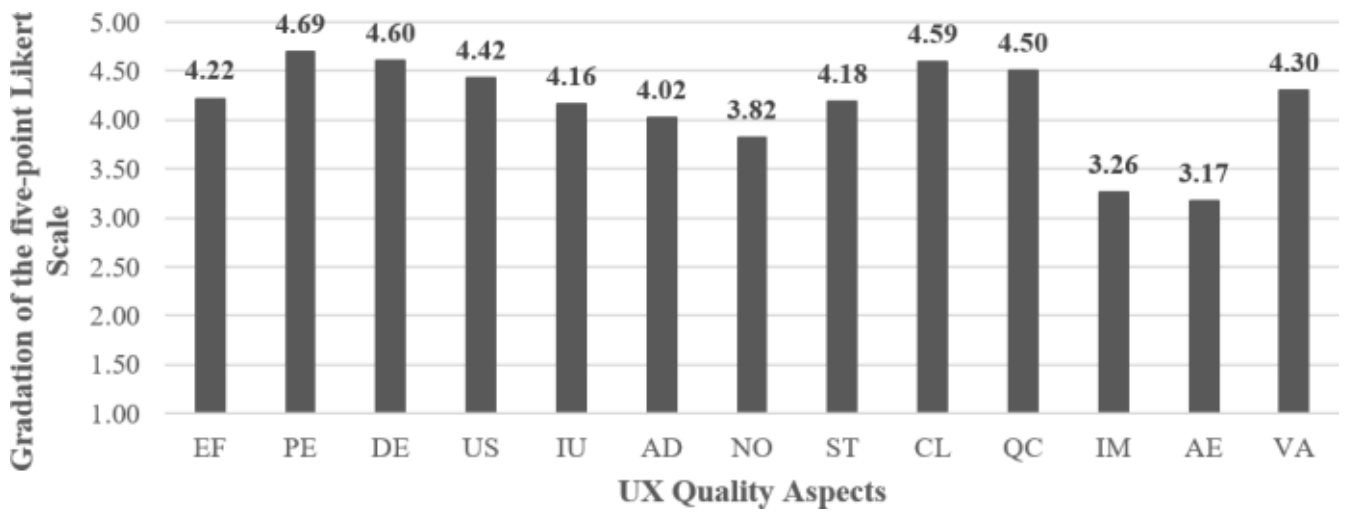

Figure 1. Importance Rating of UX Quality Aspects Concerning AR in CT.

### A. *Determining Relevant UX Quality Aspects*

The questionnaire is based on the perspective concerning UX quality aspects according to [14]. As this is a common approach of several questionnaires, we also rely on this understanding (See Section II-A, Table I). Therefore, the relevant UX quality aspects for AR in CT were considered. Data concerning the importance of different UX quality aspects were collected over an online survey. In the following, we only refer to the main results of this survey. Details are described in our previous paper [10].

Participants could start the survey by clicking a link in an invitation mail. In an introductory part, the technology AR was explained in detail. To give participants more context, a video showing a CT scenario using AR was displayed. The 13 of the 16 UX aspects described in Table I are displayed. The aspects *Identity*, *Loyalty*, and *Trust* were excluded since they obviously play no role in the CT scenario). Participants were asked to rate the importance of each presented UX quality aspect in relation to the shown CT scenario. For the rating, we applied a five-point Likert scale (from *"not important at all"* scored as 1 to *"very important"* scored as 5). We collected 121 complete responses using this survey.

Figure 1 illustrates the overall importance ratings for the UX aspects. The Y-axis presents the gradation of the applied five-point Likert scale (from *"not important at all"–1* to *"very important"–5*). The X-axis shows the mean values for the evaluated UX quality aspects. The highest rated and thus most important aspect for AR in CT is *Perspicuity* with an average score of 4.69, whereas the lowest rated one is *Aesthetics* with an average score of 3.17. Nevertheless, none of the quality aspects were rated as unimportant (or less than 3) on average.

### B. *Construction of Measurement Items*

We decided to consider only the five most important UX quality aspects for the questionnaire to keep the length manageable. For the construction of the item pool, a large set of items from 60 established UX questionnaires (with overall around 1500 single items) was analyzed. Ten suitable items for each UX quality aspect were extracted and reformulated to match our research context of mobile augmented reality in learning. This initial list of statements was then again reviewed concerning their formulations and potential duplicates or items that were, after the reformulation, too similar to other items and were removed. In the next step, the study's three authors independently selected the most representative items based on their expertise.

This resulted in the following list of candidate items (the term in brackets is later used to refer to the item):

**Efficiency**

- Using the application for learning is practical (EF1)
- The application reduces the learning effort (EF2)
- The application helps me to learn faster (EF3)
- The application saves me time while learning (EF4)
- The application improves my learning and work performance (EF5)

**Perspicuity**

- It was clear from the start how I had to use the application for learning (PE1)
- It is easy/simple to learn how to use the application (PE2)
- The information in the application is easy to understand (PE3)
- The operation of the application is logical (PE4)
- It is easy to navigate between individual parts of the application (PE5)

**Dependability**

- The behavior of the application always meets my expectations (DE1)
- I am confident in using the application at all times (DE2)
- The application is easy to control (DE3)
- I always have control over the application at every step (DE4)
- It is easy to find your way around the application (DE5)
- The application always responds comprehensible (DE6)

**Usefulness**

- The application helps me to learn (US1)
- It is a great advantage to use the application when learning (US2)
- The application is useful for learning (US3)
- I find the application useful for learning (US4)
- The application fully meets my expectations (US5)

**Clarity**

- The information on the display is clearly laid out (CL1)
- The information on the display is clear (CL2)
- The display of the application looks tidy (CL3)
- It's easy to find the information I need (CL4)

We collected data in a study with German participants. Thus, the items were carefully translated into German. The German items can be found in the appendix. A seven-point Likert scale was applied (See Figure 2). We chose an emoji-based scale to reduce mental effort and increase both attention and clarity for the trainees.

The following describes the evaluation scenario and procedure of the empirical study.

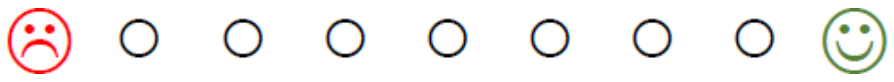

Figure 2. Emoji-based seven-point Likert Scale.

## IV. Empirical Study Design for the Questionnaire Evaluation

We conducted an empirical study to evaluate the constructed UXAR-CT questionnaire within AR-based CT learning scenarios. In the following, we present the respective AR-based CT scenarios and the study procedure.

### A. *Evaluation Scenario: Augmented Reality-based CT Applications*

We collaborated with the Chamber of Crafts for Lower Franconia in Schweinfurt, Bavaria (Germany) for the evaluation and data collection. In summary, 53 Chambers of Crafts can be found in Germany. The main concept is representing the entire crafts sector in Germany. Moreover, they are responsible for the education and training of apprentices in the craft sector, including over 130 apprenticeships in the fields of construction, wood, metal/electrical, clothing, food, health, glass, and paper. Therefore, various courses take place at the chambers, which are relevant and mandatory for the apprenticeships [28].

The institution in Schweinfurt has carried out the project **ARihA – Augmented Reality in Corporate Training**. The idea of the project was to develop and implement innovative digital learning and teaching methods using AR in CT, resulting in an immersive and action-oriented learning experience. The aim was to enhance the learning effectiveness and motivation of the trainees. The ARihA project was funded by the German Federal Ministry of Education and Research (BMBF). In the context of the project, five AR-based CT scenarios among three craft sectors were developed. The scenarios are part of the apprenticeship in the electrical engineering, metal construction, and automotive engineering sectors. Head-mounted displays (Holo-Lens 2) and handheld devices (tablets) were applied for the applications. An overview is given in the following Table II [29]. Moreover, some examples from the AR-based CT scenarios are shown in Figures 3 and 4.

TABLE II. AR APPLICATIONS FOR CORPORATE TRAINING [14]

| Training Process | Craft Sector | AR Device |
|---|---|---|
| Testing of electronic devices | electrical engineering | head-mounted |
| Processing of high-grade steel and aluminum | metal construction | head-mounted |
| Installation of locking and security systems | metal construction | handheld |
| Troubleshooting and use of measurement devices on a car lighting wall | automotive engineering | both |
| Changing the timing belt on a car engine | automotive engineering | handheld |

Further information can be found online [29]. The described AR applications of the CT scenarios are the basis for evaluating the UXAR-CT. The study approach is described below.

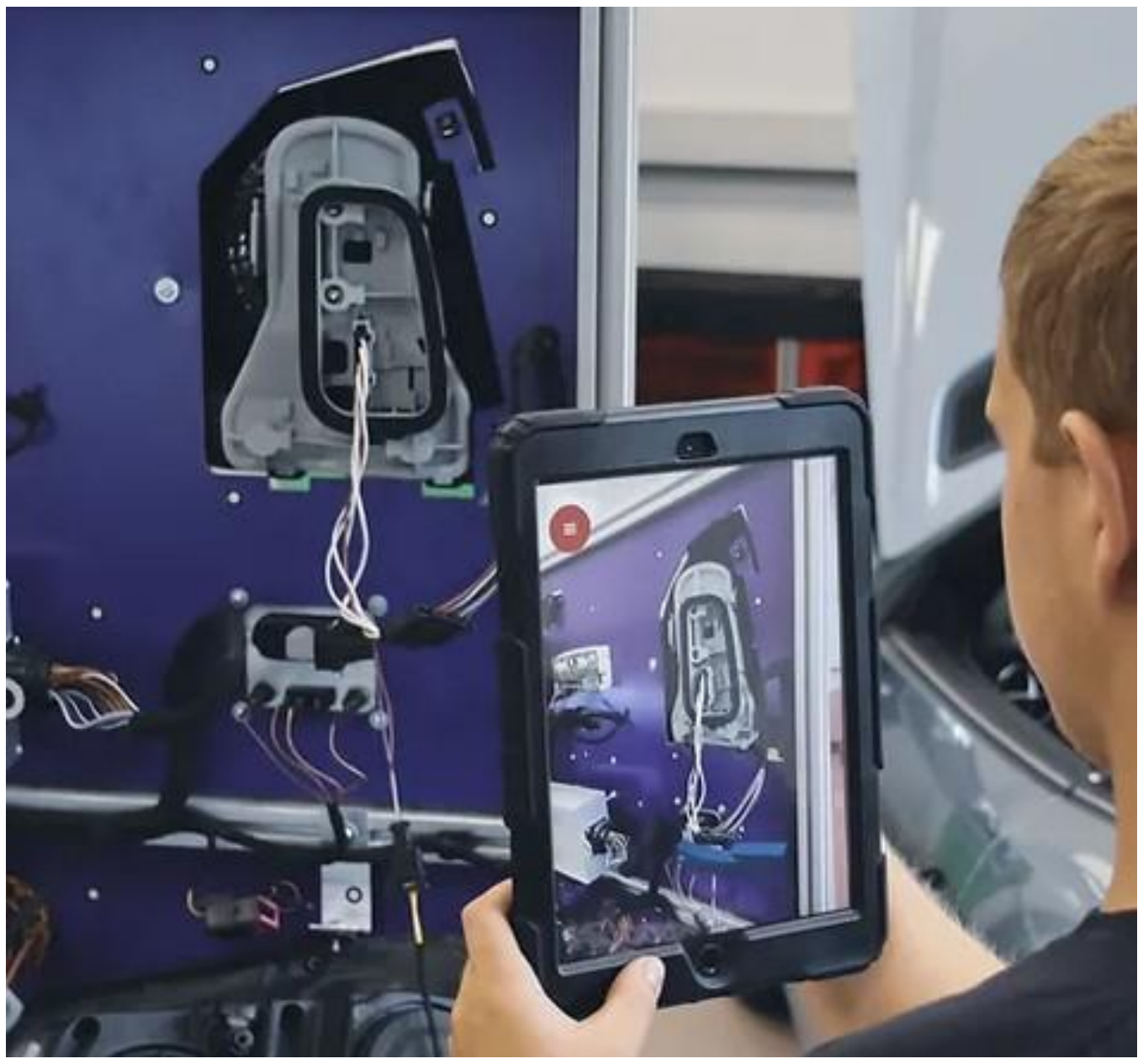

Figure 3. Troubleshooting and use of measurement devices on a car lighting wall (1) [10][29].

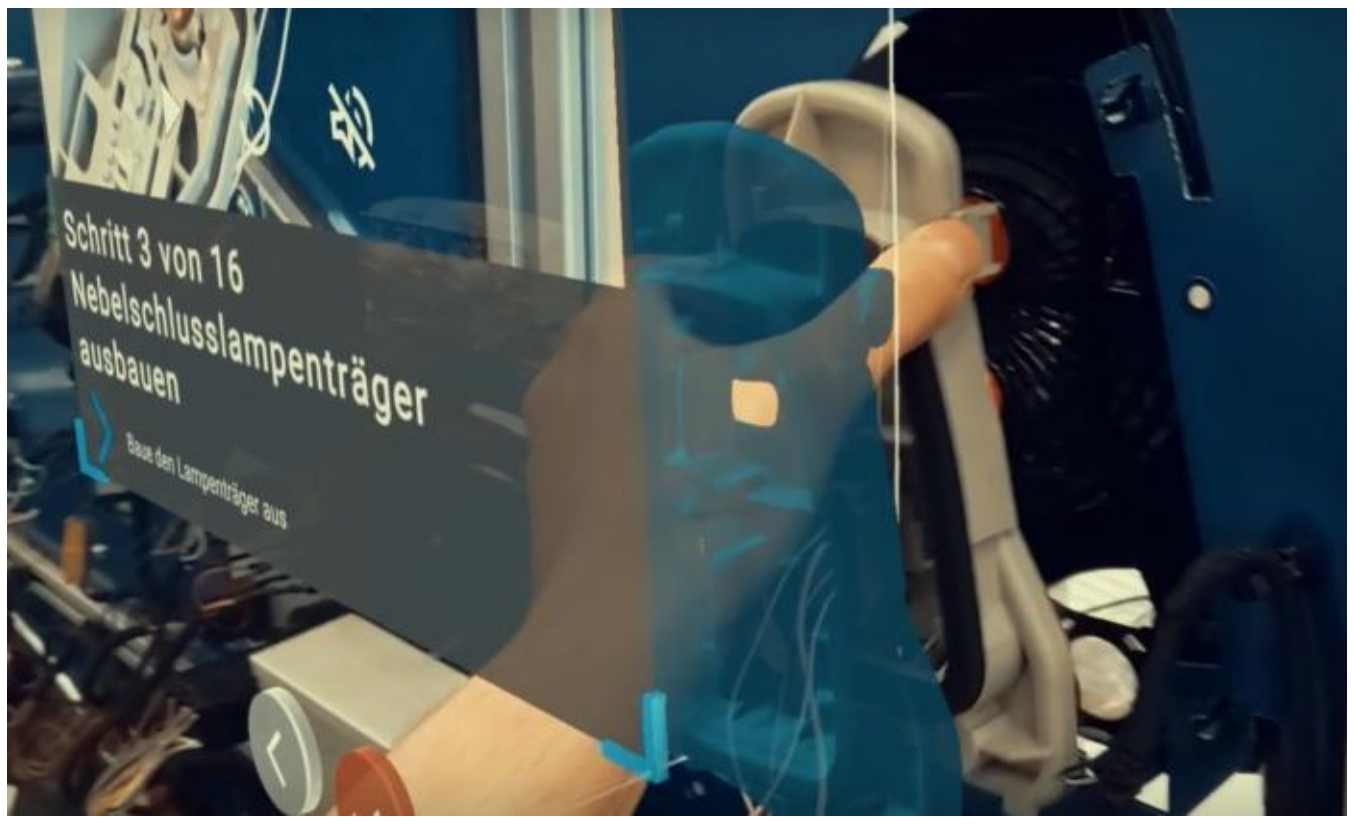

Figure 4. Troubleshooting and use of measurement devices on a car lighting wall (2) [10][29].

### B. *Empirical Survey Structure and Procedure*

We applied the UXAR-CT questionnaire to the five AR-based CT scenarios in the Chamber of Crafts. The AR applications of the different CT scenarios are regularly used in daily teaching activities. After the participants finished their learning activities, they filled out a survey (paper-pencil) containing a short motivation and instruction.

The instruction was followed by questions concerning age, gender, the learning scenario just completed, and the participants' apprenticeship. Moreover, we added two open questions:

- *What did you like about the application?*
- *What should be improved?*

After these initial questions, the candidate items described above (See Section III-B) were presented. In addition to the candidate items, one additional question was added:

*Overall, I am satisfied with the support provided by the application for my learning tasks*.

This is a classical item to measure overall satisfaction, and the responses should help to select the best-fitting items for a scale (see description below). To sum up, the applied UXAR-CT questionnaire has the following structure:

1) Motivation and Instruction
2) Demographics (4 Items)
3) Open Questions (2 Items)
4) Overall Satisfaction (1 Item)
5) Efficiency (5 Items)
6) Perspicuity (5 Items)
7) Dependability (6 Items)
8) Usefulness (5 Items)
9) Clarity (4 Items)

The items referring to the UX scales (See Section III-B) were presented in random order. The questionnaire contains 32 items, 26 of which relate to the UX (Overall Satisfaction included).

We collected responses to our survey from December 2023 to May 2024 in the Chamber of Crafts for Lower Franconia in Schweinfurt, Bavaria (Germany). The course instructors were briefed on the questionnaire and were available to answer the trainees' questions during completion. In this regard, we refer to the concept of the Chamber of Handicrafts. Courses usually last one week. In addition, courses, including AR-based CT scenarios, are not held every week. This emphasizes the difficulty of data collection in the CT application field. The evaluation results of the UXAR-CT are presented below.

# V. Evaluation Results of the UXAR-CT Questionnaire

This Section V provides the evaluation results of the empirical study. Both descriptive results and results of the Principal Component Analysis (PCA) are shown. The best-fitting items were selected. Moreover, the first results regarding scale quality are presented. Finally, further development suggestions are provided.

## A. Descriptive Evaluation Results

We first want to present the descriptive evaluation results of the empirical study. During the study period of six months, a total of ***106*** completed questionnaires were collected. There were 12 female and 93 male participants. One person did not provide any information about the gender. The average age is 19.

Moreover, we reviewed the qualitative results regarding the open questions. Thus, some qualitative insights can be provided. The trainees were generally positive about the AR-based CT scenarios. In the following, we list the qualitative answers mentioned at least five times or more. The trainees particularly liked the visual presentation of the learning and teaching content and 3D objects. Furthermore, it was perceived as simple and supportive to understand the explanations regarding the learning content. Moreover, the interaction with the learning content was perceived as useful. Additionally, the use and handling of the applications was described as simple. In summary, using AR was perceived as novel, fun, and varied in contrast to other learning methods.

However, some aspects for improvement were declared. In this context, the reaction and function of the system has been mentioned. In some cases, this did not always work correctly. Furthermore, the quality of learning and teaching content presented was criticized. Lastly, it was noted that extended use can cause both headaches and eye pain. Moreover, the comfort of wearing HoloLens2 decreases. The following Table III summarizes the qualitative results.

TABLE III. QUALITATIVE INSIGHTS FROM THE EVALUATION.

| Positive Aspects | Improvement Potential |
|---|---|
| visual representation and 3D objects | content quality |
| simplicity and support of understanding | functionality and system reaction |
| interaction | headaches and eye pain |
| simple handling | wearing comfort |
| novelty, fun, and variety | |

## B. Semantic Homogenity

A Principal Component Analysis (PCA) [30][31] with varimax rotation was performed for each of the proposed scales. The goal of this analysis was first to see if the candidate items are uni-dimensional or if they split into separate components that represent different semantic meanings. Secondly, the loading of the single items on the corresponding component demonstrates how well the items represent this component.

We show in the following the results of the PCA per suggested scale. The figures show the eigenvalues of the components. To determine if a scale is semantically homogeneous, we use the scree-test [32] (determine the point after which eigenvalues differ only slightly) and the Kaiser-Gutmann [33] criterion (remove components with eigenvalues less than 1).

***Efficiency***: The scree test and the Kaiser-Gutmann criterion (See Figure 5) both indicate a solution with one single component. This component explains 76% of the variability in the data.

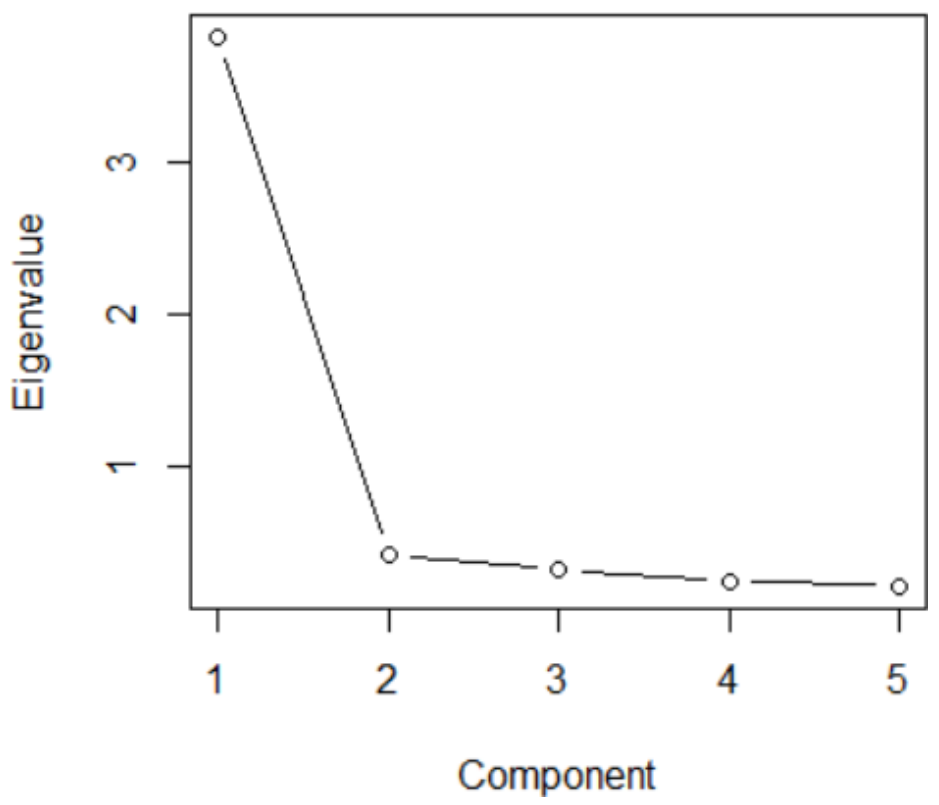

Figure 5. Screen plot of the eigenvalues for the efficiency scale.

***Perspicuity***: Again, both criteria (See Figure 6) indicate a solution with one single component that explains 64% of the variability in the data.

***Dependability***: Also, for this semantic group of items, both criteria (See Figure 7) indicate a single component. The component explains 66% of the variability in the data.

***Usefulness***: The scree test and the Kaiser-Gutmann criterion (See Figure 8) both indicate a solution with one single component. This component explains 79% of the variability in the data.

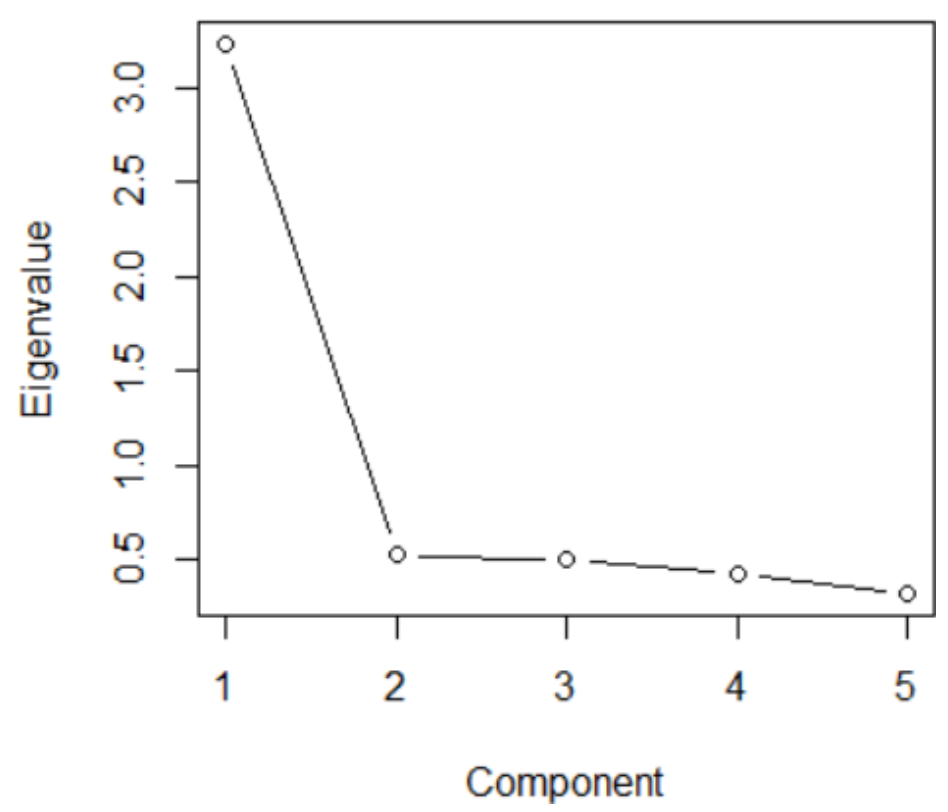

Figure 6. Screen plot of the eigenvalues for the perspicuity scale.

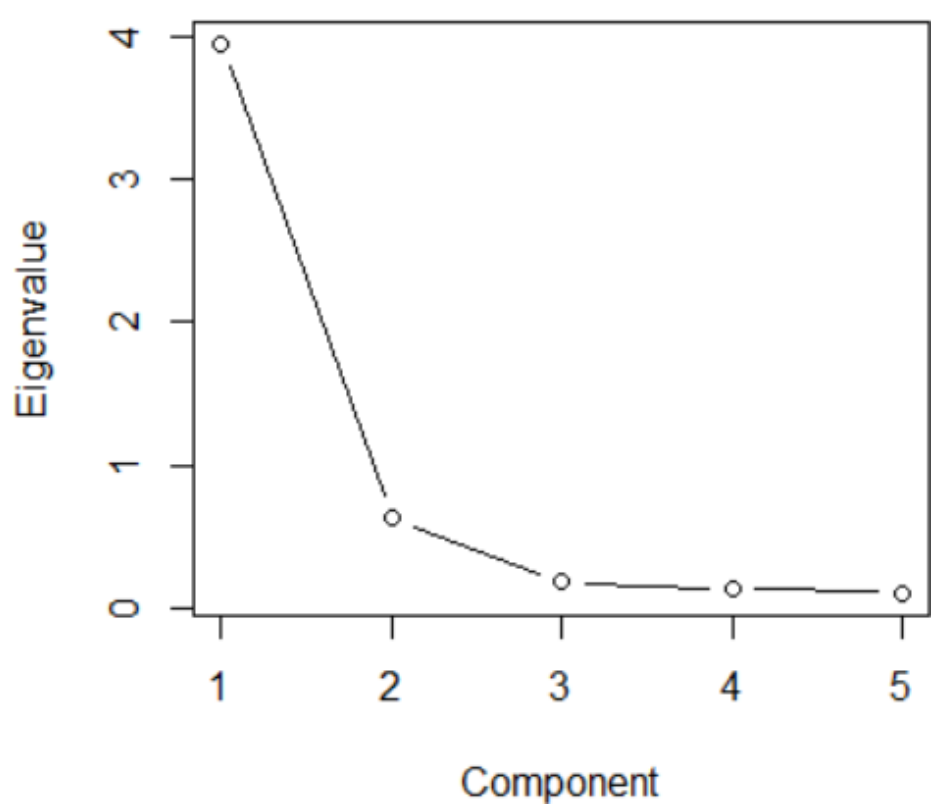

Figure 8. Screen plot of the eigenvalues for the usefulness scale.

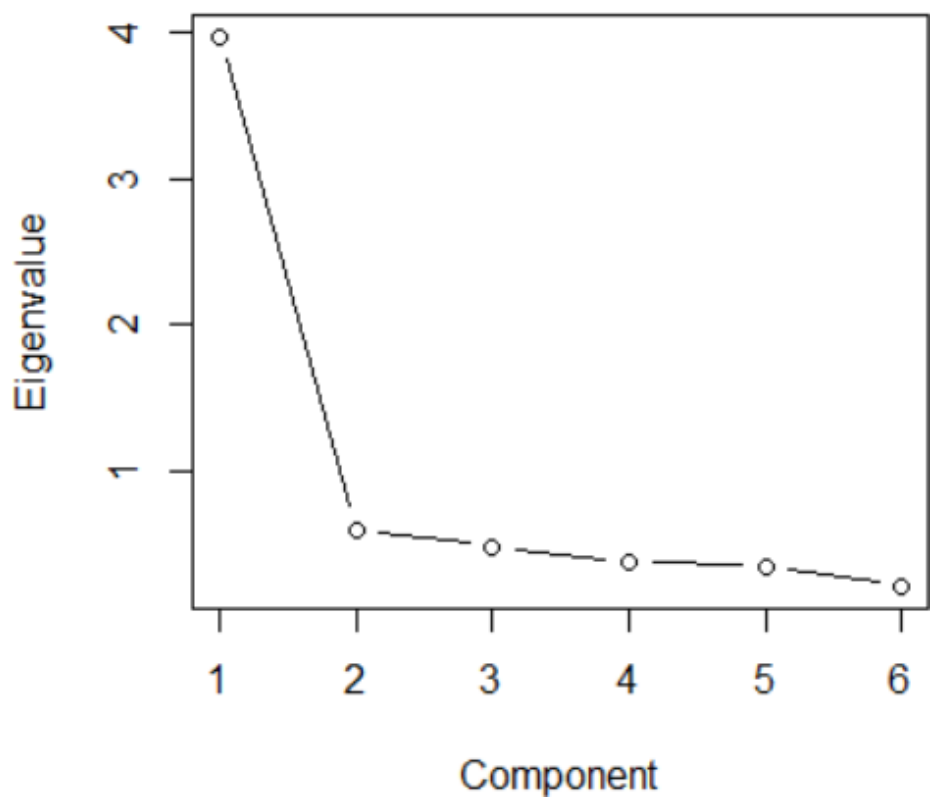

Figure 7. Screen plot of the eigenvalues for the dependability scale.

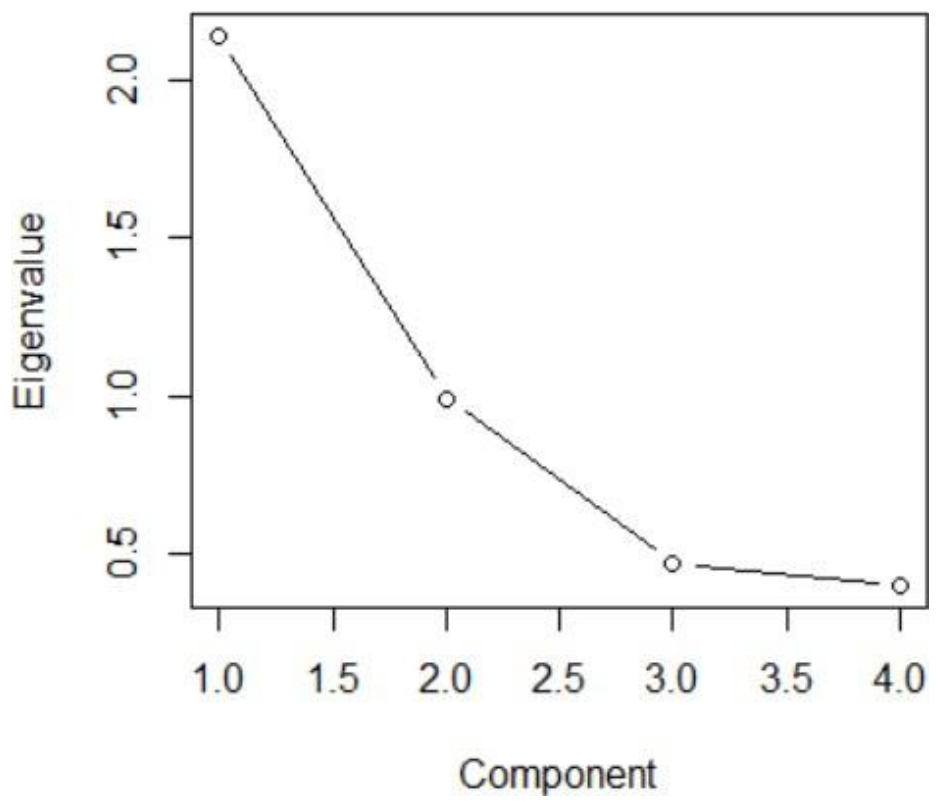

Figure 9. Screen plot of the eigenvalues for the clarity scale.

***Clarity***: Again, the scree test and the Kaiser-Gutmann criterion (See Figure 9) indicate considering only a single component. This component explains 53% of the variability in the data.

Thus, the analysis indicates semantic homogeneity of the candidate items for all proposed scales. Therefore, we can safely assume that all candidate items in a proposed scale measure the same semantic concept and can now, in the next step, identify those items that form the best representation of this concept.

## C. *Selection of the Items*

We use established criteria to select the best-fitting items per scale. Firstly, guidelines suggested by [34] state that loadings greater than 0.4 are generally considered acceptable. Thus, items with a lower loading are removed from the selection. Secondly, the higher the loading, the better the component represented by the item. However, in cases where the loadings of the remaining items are very similar, we used the item’s correlation to the overall satisfaction rating as a basis for decision-making.

For Efficiency, all items show very similar loadings on the first component. The items EF1 (0.72), EF3 (0.62), and EF5 (0.62) show clearly higher correlations to the overall satisfaction than EF2 (0.41) and EF4 (0.47) and are thus selected to represent this scale (values in parenthesis are the correlations). For Perspicuity, all items show similar loadings on the component. Based on the correlations to overall satisfaction, the items PE2 (0.63), PE3 (0.63), and PE4 (0.55) were selected to represent the scale since they show a slightly higher correlation to overall satisfaction than PE1 (0.53) and PE5 (0.51), but the difference is not massive in this case. For Dependability, items DE3 (0.63), DE4 (0.55), and DE5 (0.51) showed higher loadings on the component than DE1, DE2, and DE6 and are thus selected. For Usefulness, items US1 (0.67), US2 (0.67), US3 (0.65), and US4 (0.68) showed similar loadings on the component and similar correlations with the overall satisfaction, while the loading of US5 was clearly lower. We select US1, US2, and US4 to represent this scale. For Clarity, item CL3 shows a much lower loading than the other three items, CL1 (0.54), CL2 (0.48), and CL4 (0.55), which are therefore selected.

The final version of the UXAR-CT in English can be found in Appendix B.

In the following, preliminary results concerning scale quality are illustrated.

## D. *First Results concerning Scale Quality*

From our data set, we can calculate the Cronbach Alpha coefficient, which is a basic score for internal scale consistency. The corresponding values are 0.90 for Efficiency, 0.81 for Perspicuity, 0.85 for Dependability, 0.95 for Usefulness, and 0.79 for Clarity. These scores indicate a high level of internal consistency. Further quality criteria must be determined in the practical application of the questionnaire.

### E. Further Development Suggestions

Based on the results, we can present a reliable questionnaire with the UXAR-CT. However, the questionnaire indicates potential for further development. On the one hand, there is no weighting along the five UX scales regarding importance and relevance. On the other hand, the UXAR-CT only provides a purely UX perspective. Both aspects are discussed in the following.

When analyzing questionnaires with numerical evaluation results, the question of how these can be interpreted always arises. It is common practice to compare the results with others. However, there is a lack of common ground within the UX questionnaires differing in structure and focus [5][8][9][15]. In addition, the UXAR-CT is a domain-specific questionnaire. This results in the difficulty of comparing evaluation results. Establishing a benchmark, such as with the UEQ, is common practice. However, this requires a large amount of data [19][35]. Creating a benchmark for a domain-specific questionnaire like the UXAR-CT is almost impossible. Another way to differentiate between the scales is by including and using an ***external criterion***. Thus, the relationship between the external criterion and the scales based on their correlation and evaluation results can be considered, and weightings can be derived. Concerning the UXAR-CT, we applied the *Overall satisfaction* as an external criterion. Thus, we can use this for further development to determine the relevance of the different UX scales.

Another aspect are the the questionnaire components. Up to now, the UXAR-CT only contains UX quality aspects. While these quality aspects were identified as relevant for AR in CT (See Section III-A, [10]), there is no relation to specific system properties concerning the applications. In other words, a developer can only draw limited conclusions about the improvement potential of the system properties on the basis of the evaluation results. Relevant measurement items referring to specific AR system properties should be integrated. In this way, practical suggestions for improving the AR-based CT applications could be implemented, and thus, the learning experience could be further improved. In this regard, we refer to an exemplary questionnaire with scales and items addressing relevant system properties: the *ARcis questionnaire*. The ARcis conceptualizes the three AR characteristics by [1] and relates them to the learning context. The questionnaire measures learners' perceptions of these characteristics [36].

## VI. Conclusion

This article is based on our previous research [10] and describes the final construction of the UXAR-CT, a domain-specific standardized UX questionnaire referring to AR-based CT scenarios. We conducted an empirical survey applying the first version of the UXAR-CT to evaluate different AR-based CT applications at the Chamber of Handicrafts for Lower Franconia in Schweinfurt (Germany). We determined the best-fitting items per scale by performing a PCA based on the evaluation results. As a result, we provide a reliable questionnaire based on established UX quality aspects from a user perspective.

### A. Implications

No standardized UX questionnaire for AR exists in the CT application field. Thus, there is a lack of research. By constructing the UXAR-CT questionnaire, we provide a valuable contribution to this research field. The UXAR-CT, therefore, is based on an established common ground referring to UX quality aspects by [14]. Based on this, the importance of the respective quality aspects in relation to AR in CT was evaluated. Therefore, the questionnaire consists of UX quality aspects that are relevant to AR in CT. Thus, we can provide a reliable questionnaire based on the relevant theoretical foundation from a research perspective. The questionnaire can be applied in AR-based CT scenarios to evaluate the subjective impression of the users in practical settings. This provides valuable insights into the AR applications to improve them further.

### B. Outlook and Future Research

Finally, we want to give an outlook for future research activities. The UXAR-CT should be regularly applied in different AR-based CT scenarios in different organizations to gather further data for validation, which is the last step of a questionnaire creation process. Moreover, the questionnaire was constructed in German. Thus, an English version of the UXAR-CT would be useful for further expansion of the questionnaire and the data collection process. Additionally, we declared some suggestions for further development (See Section V-E. Up to now, the UXAR-CT only covers UX quality aspects as components. In future research activities, we want to expand the questionnaire to include relevant measurement items concerning AR system properties, enabling practical derivations regarding AR application improvement.

## Acknowledgement

We want to express our extraordinary thanks to our research partner, the Chamber of Crafts for Lower Franconia in Schweinfurt, Bavaria (Germany), for supporting this research.

# Appendix

## A) German Translations of the Measurement Items

### Overall satisfaction (Gesamtzufriedenheit)

- Insgesamt bin ich mit der Unterstützung der Anwendung fürs Lernen zufrieden.

### Efficiency (Effizienz)

- Die Nutzung der Anwendung fürs Lernen ist praktisch. (EF1)
- Die Anwendung reduziert den Lernaufwand. (EF2)
- Die Anwendung hilft mir schneller zu lernen. (EF3)
- Die Anwendung spart mir Zeit beim Lernen. (EF4)
- Die Anwendung verbessert meine Lern- und Arbeitsleistung. (EF5)

### Perspicuity (Durchschaubarkeit)

- Es war von Beginn an klar, wie ich die Anwendung zum Lernen nutzen muss. (PE1)
- Es ist leicht/einfach zu lernen, wie man die Anwendung benutzt. (PE2)
- Die Informationen der Anwendung sind einfach zu verstehen. (PE3)
- Die Bedienung der Anwendung ist logisch. (PE4)
- Es ist einfach, zwischen einzelnen Teilen der Anwendung zu navigieren. (PE5)

### Dependability (Steuerbarkeit)

- Das Verhalten der Anwendung entspricht stets meinen Erwartungen. (DE1)
- Ich bin zu jeder Zeit sicher im Umgang mit der Anwendung. (DE2)
- Die Anwendung ist einfach zu kontrollieren. (DE3)
- Ich habe bei allen Schritten immer die Kontrolle über die Anwendung. (DE4)
- Es ist einfach sich in der Anwendung zurechtzufinden. (DE5)
- Die Anwendung reagiert immer verständlich. (DE6)

### Usefulness (Nützlichkeit)

- Die Anwendung hilft mir beim Lernen. (US1)
- Es ist von großem Vorteil, die Anwendung beim Lernen zu nutzen. (US2)
- Die Anwendung ist nützlich beim Lernen. (US3)
- Ich finde die Anwendung fürs Lernen nützlich. (US4)
- Die Anwendung erfüllt meine Erwartungen vollständig. (US5)

### Clarity (Übersichtlichkeit)

- Die Darstellung der Informationen im Display ist übersichtlich. (CL1)
- Die Darstellung der Informationen im Display ist klar. (CL2)
- Das Display der Anwendung wirkt aufgeräumt. (CL3)
- Es ist einfach, die Informationen zu finden, die ich benötige. (CL4)

## B) Final UXAR-CT Questionnaire

**UXAR-CT Questionnaire**

| | |
|---|---|
| Overall, I am satisfied with the support provided by the application for my learning tasks. | ☹ ○ ○ ○ ○ ○ ○ ○ ☺ |
| Using the application for learning is practical. | ☹ ○ ○ ○ ○ ○ ○ ○ ☺ |
| The application helps me to learn faster. | ☹ ○ ○ ○ ○ ○ ○ ○ ☺ |
| The application improves my learning and work performance. | ☹ ○ ○ ○ ○ ○ ○ ○ ☺ |
| It is easy/simple to learn how to use the application. | ☹ ○ ○ ○ ○ ○ ○ ○ ☺ |
| The information in the application is easy to understand. | ☹ ○ ○ ○ ○ ○ ○ ○ ☺ |
| The operation of the application is logical. | ☹ ○ ○ ○ ○ ○ ○ ○ ☺ |
| The application is easy to control. | ☹ ○ ○ ○ ○ ○ ○ ○ ☺ |
| I always have control over the application at every step. | ☹ ○ ○ ○ ○ ○ ○ ○ ☺ |
| It is easy to find your way around the application. | ☹ ○ ○ ○ ○ ○ ○ ○ ☺ |
| The application helps me to learn. | ☹ ○ ○ ○ ○ ○ ○ ○ ☺ |
| It is a great advantage to use the application when learning. | ☹ ○ ○ ○ ○ ○ ○ ○ ☺ |
| I find the application useful for learning. | ☹ ○ ○ ○ ○ ○ ○ ○ ☺ |
| The information on the display is clearly laid out. | ☹ ○ ○ ○ ○ ○ ○ ○ ☺ |
| The information on the display is clear. | ☹ ○ ○ ○ ○ ○ ○ ○ ☺ |
| It's easy to find the information I need | ☹ ○ ○ ○ ○ ○ ○ ○ ☺ |